\documentclass[aps,pre,twocolumn,amsmath,amssymb,superscriptaddress]{revtex4-1}

\usepackage{graphicx}
\usepackage{dcolumn}
\usepackage{bm}
\usepackage{hyperref}
\usepackage[usenames,dvipsnames]{color}

\newcommand{\fig}[1]{Figure~\ref{#1}}
\newcommand{\eq}[1]{Eq.~\ref{#1}}

\newcommand{\force}{f}
\newcommand{\efield}{E}
\newcommand{\fric}{\xi}
\newcommand{\mob}{\mu}
\newcommand{\mono}{\sigma}
\newcommand{\orderParam}{S}
\newcommand{\asymm}{\varphi}
\newcommand{\effCharge}{{Q_\textmd{eff}}}
\newcommand{\kbt}{k_\textmd{B}T}
\newcommand{\rf}{R_\textmd{eff}}
\newcommand{\rg}{R_{g}}
\newcommand{\rgo}{R_{g0}}
\newcommand{\rgz}{R_{gz}}
\newcommand{\rgr}{R_{gr}}
\newcommand{\z}{z}
\newcommand{\debye}{\lambda_\textmd{D}}
\newcommand{\efffric}{\fric_\textmd{eff}}
\newcommand{\rodfric}{\fric_\textmd{rod}}
\newcommand{\av}[1]{\left\langle #1 \right\rangle}

\begin{document}
\title[Electrophoretic Mobility]{Electrophoretic Mobility of Polyelectrolytes within a Confining Well}
\thanks{Accepted:\\\textmd{ACS Macro Letters}, mz-2015-000768.R1, April 8, 2015}

\author{Tyler N. Shendruk}%
\affiliation{The Rudolf Peierls Centre for Theoretical Physics, Department of Physics, Theoretical Physics, University of Oxford, 1 Keble Road, Oxford, OX1 3NP, United Kingdom}
\email{tyler.shendruk@physics.ox.ac.uk }
\homepage{ http://tnshendruk.com/}

\author{Martin Bertrand}%
\author{Gary W. Slater}
\affiliation{Department of Physics, University of Ottawa, 150 Louis-Pasteur, Ottawa, Ontario, K1N 6N5, Canada}


\begin{abstract}
We present a numerical study of polyelectrolytes electrophoresing in free solution while squeezed by an axisymmetric confinement force transverse to their net displacement. 
Hybrid multi-particle collision dynamics and molecular dynamics simulations with mean-field finite Debye layers show that even though the polyelectrolyte chains remain ``free-draining'', their electrophoretic mobility increases with confinement in nanoconfining potential wells. 
The primary mechanism leading to the increase in mobility above the free-solution value, despite long-range hydrodynamic screening by counterion layers, is the orientation of polymer segments within Debye layers. 
The observed length-dependence of the electrophoretic mobility arises due to secondary effects of counterion condensation related to confinement compactification. 
\end{abstract}

\maketitle

Understanding the electrophoresis of confined polyelectrolytes is crucial to the advancement of many separation techniques~\cite{dorfman13}. 
Methods such as translocation through nanopores~\cite{dehaan13,laohakunakorn13,rowghanian13} and sieving through arrays of microscopic posts~\cite{randall04,randall06,dorfman10,viero11,park12} belong to a family of techniques that depend on nanoengineered geometrical constraints. 
In particular, the motion of polyelectrolytes (such as DNA) in narrow nanochannels raises a number of fundamental questions that are not yet fully understood in spite of practical interest~\cite{mannion2006cas,reisner2005sad}. 

It is well-known that long, electrophoresing polyelectrolyte chains are ``free-draining'': 
their behavior is described by local effective properties, with long-ranged hydrodynamic coupling mostly screened~\cite{shendruk12}. 
When an electric field $\vec{\efield}$ applies a force to a chain segment, it also applies an equal and opposite force to the diffuse layer of counterions over the characteristic Debye length
$\debye$. 
Thus, the viscous forces on the surrounding fluid are effectively cancelled with only a rapidly decaying residual hydrodynamic field beyond $\debye$~\cite{long01}. 
This causes the effective friction coefficient $\efffric$ to increase linearly with degree of polymerization $N$, just like the charge $\effCharge$; the free-solution electrophoretic mobility $\mob_0=\effCharge/\efffric$ is thus independent of chain length~\cite{viovy00}. 

\begin{figure}[tb]
 \begin{center}
  \includegraphics[width=0.5\textwidth]{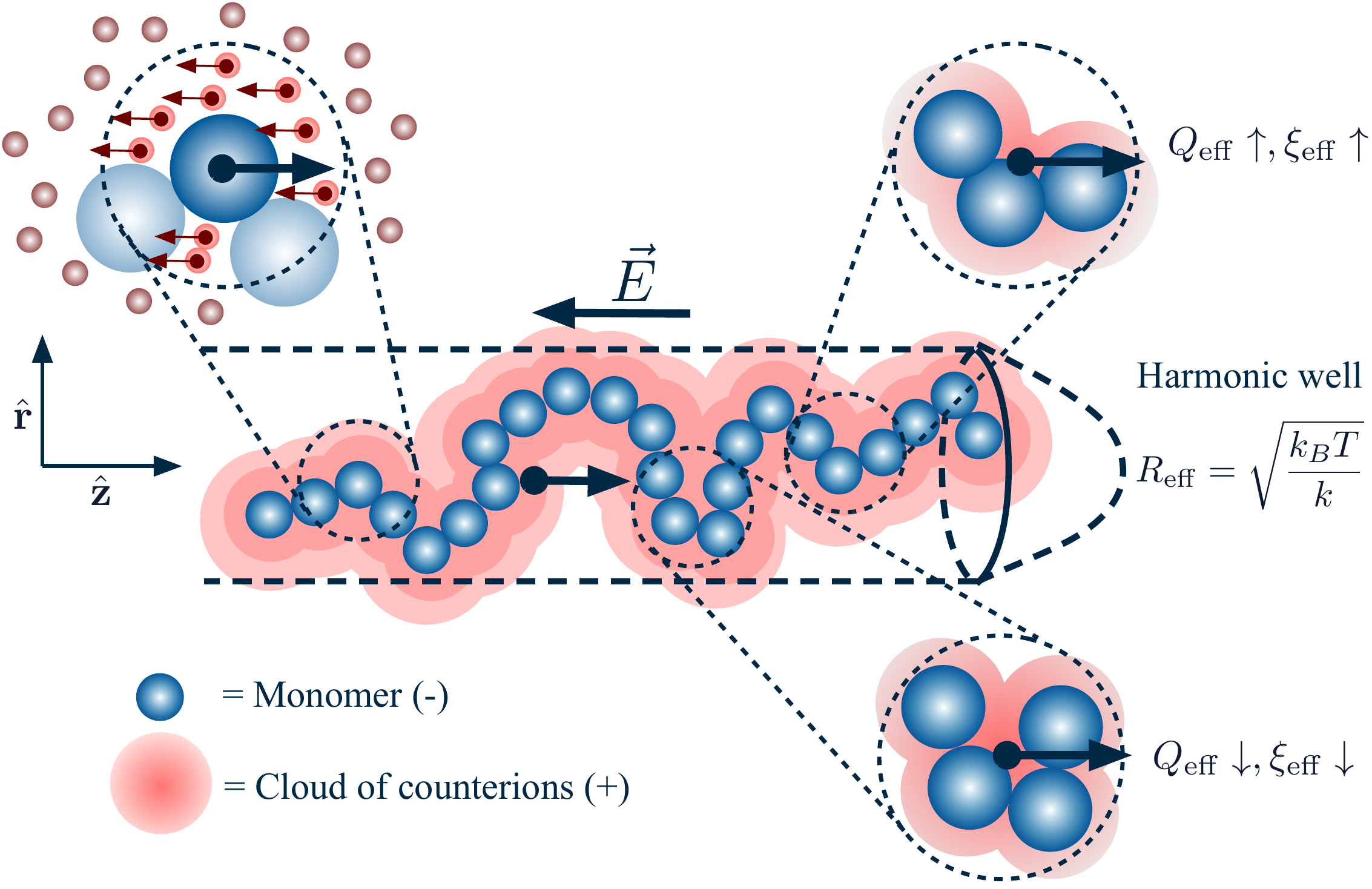}
  \caption{\small{
  A negatively charged polyelectrolyte chain undergoing electrophoretic motion under the action of an electric field $\vec{\efield}$. 
  Point-like MPCD fluid particles are assigned a charge specified by the Debye-H\"{u}ckel approximation and feel an electric force in the opposite direction to the monomer. 
  A cylindrical harmonic potential radially confines the chain, creating an effective tube of radius $\rf \equiv \sqrt{ \kbt/k }$. 
  }}
  \label{fig:sys}
 \end{center}
\end{figure}

One way to subvert this size independence is to apply a parallel mechanical force $\vec{\force}$ simultaneously with the electric field. 
If the mechanical force acts only on the monomers and not on the counterions then the net force on the fluid does not cancel, long-ranged flows are possible and the chain is no longer free-draining~\cite{desruisseaux01}. 
This concept has been applied to many situations~\cite{shendruk12} including tethered polyelectrolytes~\cite{long1996sae} and end-labeled free-solution electrophoresis~\cite{chubynsky14}. 
One oft-given example is the electrophoretic motion of a deformed polyelectrolyte through a nanofluidic channel~\cite{tlusty06,balducci06,scalieb08}: It is argued that confinement increases the frequency of collisions between the monomers and the channel walls, thereby inducing non-electric surface friction and reducing mobility~\cite{cross07}. 
However, conflicting experimental results have been reported in which mobility is observed to increase with confinement~\cite{campbell04,scalieb08,salieb2009electrokinetic}.

In order to resolve this discrepancy, we explicitly study the validity of the free-draining assumption for a freely-jointed charged chain that is squeezed by a radial potential that acts transverse to an electric field $\vec{\efield}=\efield \hat{\z}$ (\fig{fig:sys}). 
Our study does away with impermeable walls that add complications by introducing shear stresses, modifying friction coefficients, generating electro-osmotic flows and screening hydrodynamic interactions~\cite{kang2006e,mathe07,dai13}. 
Only the effects of confinement on the polymer conformations are preserved. 
Unless otherwise stated, we use a harmonic potential $U_{2}=k r^2 / 2$ , where $k$ is the confinement strength and $r$ is the distance from the $\hat{z}$-axis. 
The radial potential forms a tube of effective radius $\rf \equiv \sqrt{ \kbt/k }$ that imposes lateral constraints on the polymer conformation but does not act on the fluid. 
Although the potential confines the chain, no net axial force acts on the squeezed polyelectrolyte. 
Hence, the electro-hydrodynamic equivalence principle~\cite{long1996sae,long1996emc,grosberg10} might lead us to expect the drift velocity to match the free-draining value $\mob_0\efield$. 

\begin{figure}[tb]
 \begin{center}
  \includegraphics[width=0.5\textwidth]{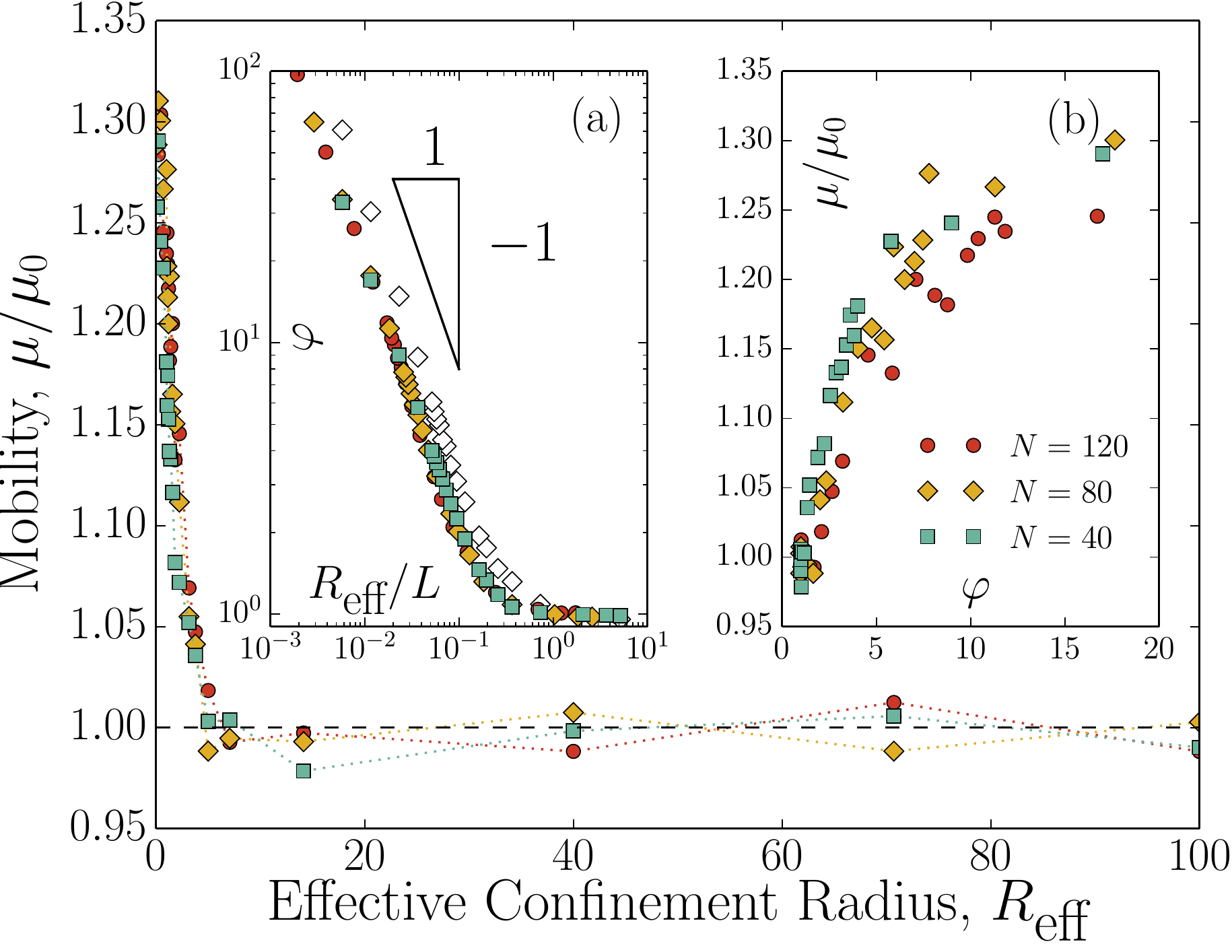}
  \caption{\small{
   Electrophoretic mobility of confined polyelectrolytes normalized by the free-solution value $\mob_{0}$ as a function of effective confinement radius $\rf=\sqrt{\kbt / k}$ for various degrees of polymerization $N$. 
   \emph{Inset (a):} 
   Asymmetry ratio $\asymm=\rgz/\rgr$ as a function of $\rf$ normalized by contour length $L$. 
   Open symbols denote neutral chains. 
   \emph{Inset (b):} 
   Mobility as a function of the asymmetry ratio $\asymm$. 
  }}
  \label{fig:mob}
 \end{center}
\end{figure}

\begin{figure}[tb]
 \begin{center}
  \includegraphics[width=0.5\textwidth]{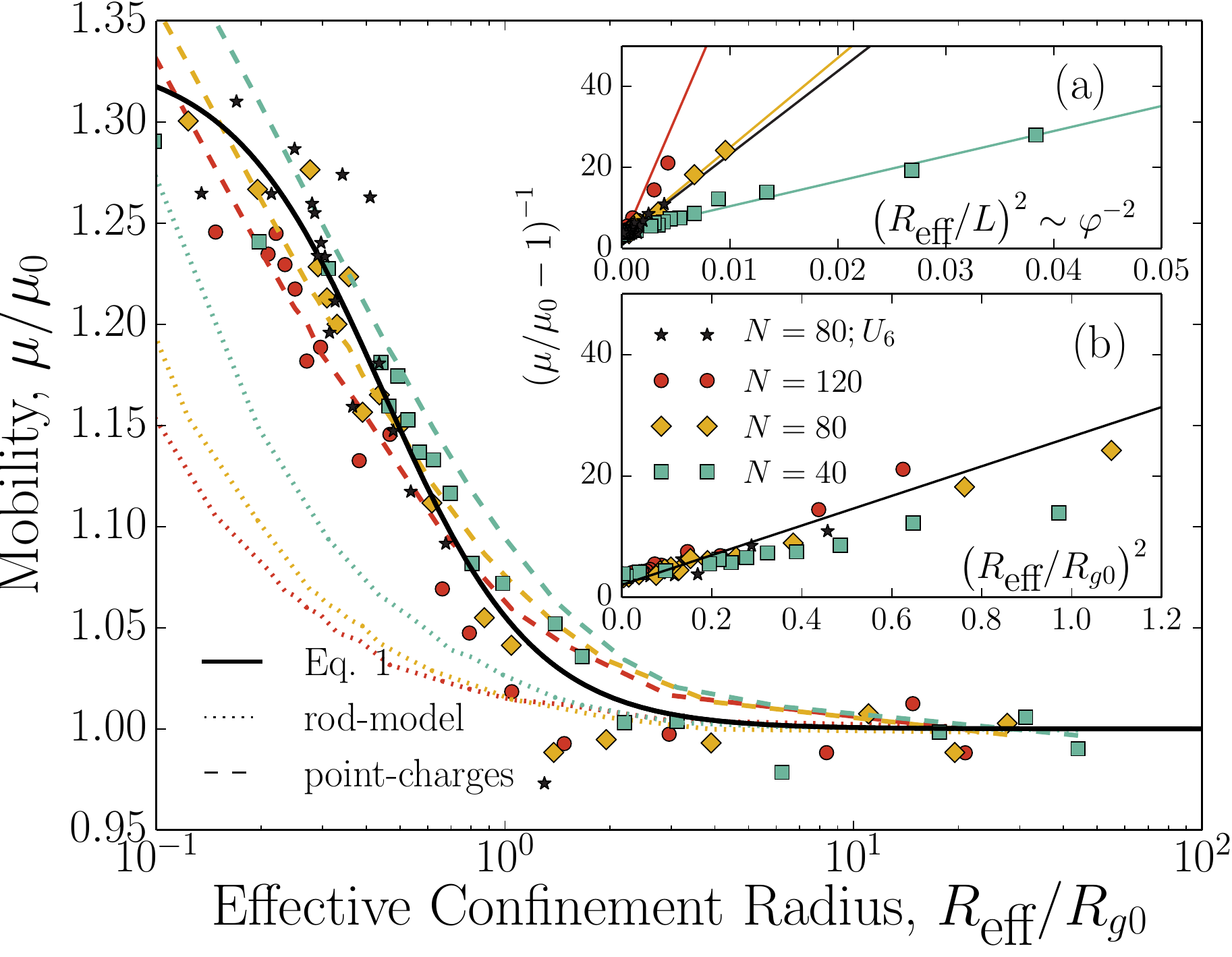}
  \caption{\small{
  Scaled mobility $\mob/\mob_{0}$ as a function of normalized effective confinement radius $\rf/\rgo$ for various degrees of polymerization $N$ and confining potentials $U_{i}$. 
  Dotted lines represent the oriented-rod model and dashed lines the Long-Ajdari screened electro-hydrodynamics model ($N=120$ scarlet; $N=80$ amber; $N=40$ beryl)
  The solid line shows \eq{eq:mob}. 
  \emph{Inset (a): } 
  Mobility dependence on $\left(\rf/L\right)^2$. 
  \emph{Inset (b): } 
  Mobility dependence on $\left(\rf/\rgo\right)^2$, consistent with the form of \eq{eq:mob}. 
  The fit produces an intercept of $3.0\pm0.1$ and slope of $15.3\pm0.3$. 
  }}
  \label{fig:compare}
 \end{center}
\end{figure}


Our simulations use the numerical methods reported by Hickey \textit{et al.}~\cite{hickey12}. 
$N$ purely repulsive Lennard-Jones charged beads (diameter $\mono$) are linked with finitely extensible non-linear elastic bonds into a polymer of contour length $L=\left(0.965\mono\right) N$. 
Monomers interact via a Debye-H\"{u}ckel potential and are embedded in a multi-particle collision dynamics (MPCD) fluid~\cite{slater09,jiang13}. 
Unless otherwise stated, the polymer chain is coupled to the MPCD fluid by including the monomers in each Andersen-MPCD collision event~\cite{noguchi07,frank09}. 
The MPCD collision cells are cubic and of size $a$, which defines the unit of length. 
We employ periodic boundary conditions on a rectangular control volume of size $35\times35\times150\;a^3$. 
MD monomers have a mass $3$ times greater than the MPCD mass scale $m$ and have one unit charge $-1e$. 
The density of the fluid is $5/a^3$, while the monomer size is $\mono=a/2$, the Debye length is $\debye=1\mono$ and the Bjerum length is $\lambda_\textmd{B}=1.5\mono$. 
The electric field is set to $\efield=1$ in simulation units~\cite{hickey12}, unless otherwise stated. 
Electro-hydrodynamic effects are implicitly simulated by assigning MPCD particles a charge based on the Debye-H\"{u}ckel approximation (\fig{fig:sys}) such that the charge density a distance $r$ away from a monomer is $\rho_e(r)\propto \exp\left(-r/\lambda_D\right)/r$.
If the resulting charge on any given fluid particle exceeds a threshold value, it is reduced for both the MPCD and the associated monomers to account for charge condensation. 
This mean-field MPCD-MD Debye-H\"{u}ckel algorithm reproduces the nonmonotonic increase in mobility with respect to $N$ of charged oligomers, as well as the nonzero mobility of certain net-neutral block polyampholytes~\cite{hickey12}. 

\begin{figure*}[tb]
 \begin{center}
  \includegraphics[trim=3.0cm 0.0cm 3.0cm 0.0cm,width=\textwidth]{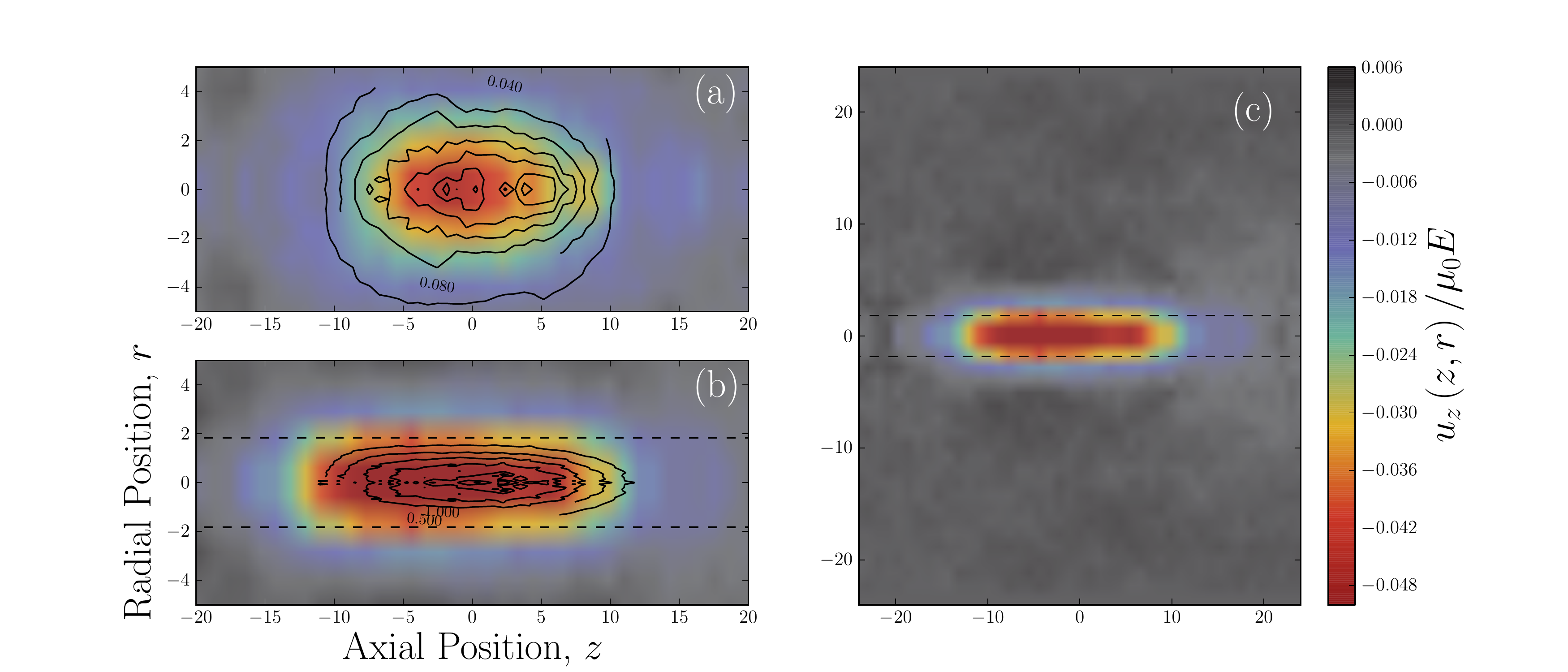}
  \caption{\small{
  Velocity of the MPCD fluid about the polyelectrolyte's centre of mass normalized by the electrophoretic velocity of the free chain. 
  The degree of polymerization is $N=200$, the electric field is $\efield=2$ and the MPCD-in-MD coupling scheme is used~\cite{padding04}. 
  \emph{Inset (a):} Fluid velocity field for 
  an unconfined polyelectrolyte. 
  \emph{Inset (b):} Velocity field for confinement by an axially symmetric harmonic potential with $k=0.3$ and an effective radius of $\rf=1.8$ (in units of MPCD collision cell size $a$). 
  \emph{Inset (c):} The far-field fluid velocity field of \fig{fig:flow}b. 
  }}
  \label{fig:flow}
 \end{center}
\end{figure*}


The conformational changes due to confinement can be characterized by the asymmetry ratio $\asymm$, which is the ratio of the axial radius of gyration $\rgz$ to the radial component $\rgr$. 
When the harmonic potential is weak ($\rf$ larger than the unconfined radius of gyration $\rgo$), the chain is unperturbed and $\asymm=1$ (\fig{fig:mob}a). 
When the effective tube size is small, the polymer deforms into a string of de~Gennes blobs~\cite{dai13b}. 
\fig{fig:mob}a shows that the asymmetry ratio of a polyelectrolyte radially confined by a harmonic potential scales as $\asymm \sim L/\rf$, as theoretically predicted for neutral chains~\cite{shew2003cbs}. 

For weak confinements, the mean-field MPCD-MD Debye-H\"{u}ckel simulations verify that the mobility is independent of effective tube diameter and chain length remaining near the free-solution value $\mob_{0}$, as expected (\fig{fig:mob}). 
However, in stronger confinements the electrophoretic mobility rises sharply above $\mob_{0}$. 
The increase in mobility above the free solution value appears roughly inversely proportional to $\rf$ in \fig{fig:mob}. 
The inset (\fig{fig:mob}b), however, reveals that this linearity is only apparently true for small asymmetries. 
Saturation to an asymptotic value of $\approx 4/3$ occurs at large $\asymm$. 

A semi-logarithmic representation of the electrophoretic mobility as a function of effective confinement radius is presented in \fig{fig:compare}. 
Also included in \fig{fig:compare} are simulations that replace $U_{2}$ with the steeper confining potential $U_{6}=kr^6/6$ that forms a tube of effective size $\rf=\sqrt[6]{\kbt/k}$ and demonstrate that the mobility rise does not depend on the harmonic nature of the confining potential. 
While \fig{fig:mob}b suggests mobility dependence on asymmetry ratio, it obscures a length dependence for a given asymmetry ratio.
\fig{fig:compare}a demonstrates that moderate axial confinement may lead to successful separation of polylectrolytes for a given asymmetry ratio. 
However, actual experiments are done at a fixed value of $\rf$, not at a fixed value of $\asymm$. Normalization by the unconfined radius of gyration $\rgo$ collapses the curves (\fig{fig:compare}b) and shows that mobility can be fit using an empirical function of the form 
\begin{align}
	\frac{\mob}{\mob_0} &\approx 1 + \frac{1}{3}\left[1+5\left(\frac{\rf}{\rgo}\right)^2\right]^{-1}. 
	\label{eq:mob}
\end{align}
We note that \eq{eq:mob} and \fig{fig:mob}b predict a plateau value $\mob/\mob_0=4/3$ for strong confinement. 
The dependence on the ratio $\left(\rf/\rgo\right)^2$ is not a result of the harmonic nature of $U_{2}$ since the proposed function (\eq{eq:mob}) remains predictive for the $U_{6}$ case as well (\fig{fig:compare}b). 
Clearly, our data and \eq{eq:mob} support the idea that longer polylectrolytes should migrate faster under confinement; however, this does not explain the physical mechanisms that leads to this potentially useful $N$-dependent rise in mobility. 

Length dependent electrophoretic mobilities often suggest that a chain is no longer free-draining~\cite{shendruk12}. 
However, we find that this is not the case --- the polyelectrolyte remains free-draining. 
If it did not, long-range electro-hydrodynamic interactions would couple distant segments and the polyelectrolyte would electrophorese as a Zimmian chain, entraining distant fluid along with it. 
We explicitly consider the entrainment of fluid
in \fig{fig:flow}. 
In the absence of confinement (\fig{fig:flow}a), a small electro-osmotic flow (magnitude $< 5\%$ of $\mob_{0} \efield$) occurs within the polymer coil. 
This is consistent with the ideal picture of a long free-draining polymer for which hydrodynamic interactions are negligible and there are no far-field perturbations to the surrounding fluid.  

The flow-field predicted by our 
simulations of an electrophoresing chain confined to an effective tube of $\rf=1.8$ is quite similar (\fig{fig:flow}b). 
If free-draining were thwarted by confinement then the entrained fluid would be expected to move with a velocity comparable to the electrophoretic velocity of the chain and long-range perturbations akin to those produced by a no-slip, impermeable body would be expected. 
However, this is not observed. 
Rather, the entrained fluid velocity remains significantly less than that of the translating chain (\fig{fig:flow}b). 
Since neither the unconfined nor the confined coils produce significant far-field flows, they can both be described as ``free-draining''. 
This is emphasized in \fig{fig:flow}c, which explicitly shows that far from the confined polyelectrolyte the velocity field is indeed zero. 

\begin{figure}[tb]
 \begin{center}
  \includegraphics[width=0.5\textwidth]{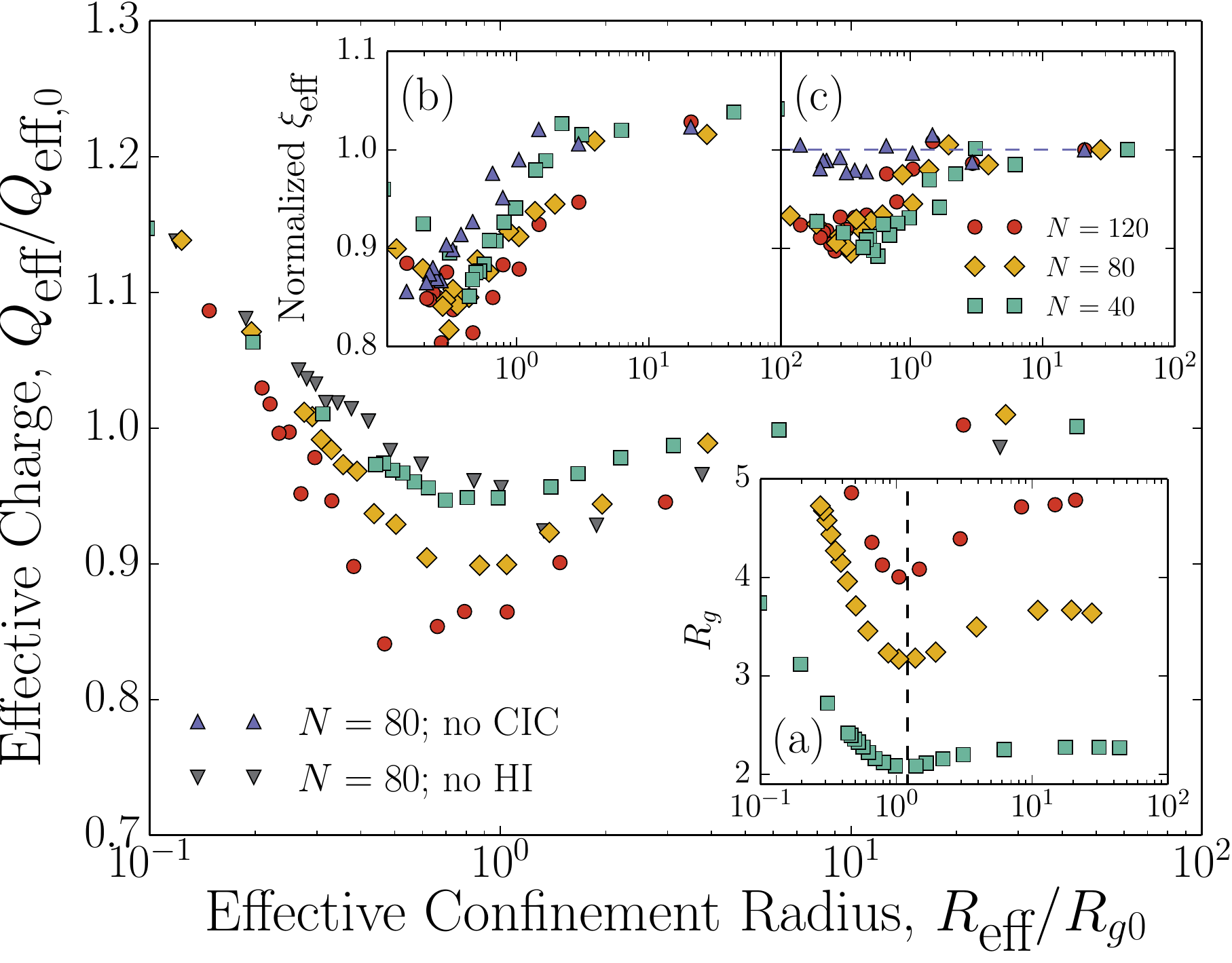}
  \caption{\small{
  The effective charge of confined polyelectrolytes (normalized by the free-solution value $\effCharge_{,0}$) as a function of normalized effective confinement radius $\rf/\rgo$ for various degrees of polymerization $N$. 
  \emph{Inset (a):} 
  Radius of gyration $\rg$ of confined chains. 
  Normalization of $\rf$ with $\rgo \sim N^{3/5}$ causes the confinement compactification minima to concur. 
  \emph{Inset (b):} The effective friction coefficient of confined chains normalized by the free-solution value ${\efffric}_{,0}$. 
  Simulations without counterion condensation (hydrodynamic interactions) are labelled ``no CIC'' (``no HI''). 
  \emph{Inset (c):} 
  The effective friction coefficient $\efffric$ normalized by the rod-model $\rodfric$, which shows the correction factor due to charge condensation. 
  }}
  \label{fig:charge}
 \end{center}
\end{figure}

Since the electrophoresing chain remains free-draining, the mobility must be a function of effective local terms. 
Because electrostatic repulsion is not fully screened within the Debye layer, local segments are locally stiff and form roughly rigid rods of length $\sim\debye$. 
In free-solution, these local segments randomly sample all orientations but when the chain is strongly confined the segments orient 
and the effective friction coefficient of the segments decreases to approximately $\efffric \approx \rodfric \approx \fric_\parallel \left( 5-2\orderParam \right) / 3$, where $\orderParam= \av{3\cos^2\theta-1}/2$ is the orientation order parameter and $\fric_\parallel$ is the friction coefficient of a slender rod oriented parallel to $\hat{z}$. 
This simple, local rod-model predicts that the electrophoretic mobility should increase as $\sim \rodfric^{-1}$, which agrees qualitatively with MPCD-MD simulations although it does not predict any $N$-dependence exhibited in simulations (\fig{fig:compare}; dotted lines). 
Likewise, more meticulously modelling of monomers as screened point-charges by the Long-Ajdari electro-hydrodynamic interaction tensor~\cite{long01} produces more accurate predictions but the mobility remains $N$-independent (\fig{fig:compare}; dashed lines). 

This leads to the conclusion that the $N$-dependence must be a result of variation of the effective charge $\effCharge$, which occurs as a higher order effect to the primary mechanism of $N$-independent friction reduction due to orientation along the tube. 
To test this explicitly, the counterion condensation is ``switched off''~\cite{hickey12} such that the mobility is determined solely by the effective friction of a segment. 
Measuring the effective friction directly from the mobility, we observe that $\efffric$ is well approximated by $\rodfric$ (\fig{fig:charge}c; no CIC). 
Thus, local friction coefficient reduction due to orientation within the Debye layer is indeed the main mechanism for the mobility increase; however, it is not sufficient to account for the $N$-dependence. 
If, instead, hydrodynamic interactions are ``switched off''~\cite{kikuchi02,kikuchi05} then the mobility still varies from its free-draining value. 
Since the friction coefficient remains constant by construction, the effective charge must vary.
\fig{fig:charge} (no HI) shows the explicitly measured $\effCharge$ as a function of confinement radius. 
An $N$-dependent minimum of $\effCharge$ exists when hydrodynamic interactions are included (\fig{fig:charge}). 
It is this $N$-dependence in $\effCharge$ that leads to the $N$-dependent mobility. 

The effective charge varies due to confinement-induced variation in monomer density and, in particular, to the $N$-dependent phenomenon of confinement compactification~\cite{micheletti11}. 
At moderate confinements, the chain first orients along $\hat{z}$ causing $\rg$ to decrease (\fig{fig:charge}a). 
Since the likelihood of finding distant segments near one another increases, diffuse layers overlap (\fig{fig:sys}) and counterion condensation increases, which lowers the effective charge relative to the unconfined coil (\fig{fig:charge}). 
More strongly confined chains are stretched into a string of blobs along the axis of the tube causing $\rg$ to increase (\fig{fig:charge}a). Blob theory predicts that the density of monomers continues to increase with confinement. However, in the strong confinement limit, the chain is highly stretched and no longer in the blob regime, such that distant segments are then less likely to be near one another. 
In turn, the fraction of condensed counterions decreases and $\effCharge$ increases toward the charged-rod limit~\cite{muthukumar04}. 
The position of the $\rg$ minimum universally depends on $\rf/\rgo \sim \rf N^{-3/5}$ (\fig{fig:charge}a)~\cite{morrison05} and, therefore, so too does $\effCharge\left(\rf/\rgo\right)$. 
It is precisely this secondary effect that causes electrophoretic mobility $\mob$ to be a universal function of the effective tube radius when normalized by unconfined radius of gyration $\left(\rf/\rgo\right)$ as seen in \fig{fig:compare}. 

Finally, counterion condensation has a tertiary effect on $\mob$. 
When more counterions condense increased coupling occurs and the effective friction coefficient decreases since the effective charge sets the number of counterions remaining in the diffuse layer screening hydrodynamic coupling. 
Likewise, when the effective charge goes up, hydrodynamic screening is enhanced and $\efffric$ rises (\fig{fig:charge}b). 
This effect ensures that the mobility is not quite $\effCharge/\rodfric$ since a correction factor $c\left(\effCharge\right) = \efffric/\rodfric$ must be accounted for in the effective drag (\fig{fig:charge}c). 
Thus, we conclude that the mobility increases from its free-draining value primarily because orientation reduces the local friction coefficient, while length dependence arises from the secondary effect of charge condensation due to confinement compactification, which in turn cause the tertiary effect of reduced hydrodynamic screening, further decreasing the effective friction coefficient. 

By considering a simplified system that confines freely-jointed polyelectrolytes via radial potentials rather than impermeable walls, this study explicitly demonstrates that the electrophoretic mobility depends on conformation in a length-dependent manner, though the chain, as a whole, remains free-draining. 
The primary effect is local alignment of segments within Debye layers, which would not be accounted for by blob theories of confined polymers that assume uniform monomer density and orientation distributions within blobs~\cite{werner13}. 
Simulations of this effect require computational techniques that account for finite Debye layers, such as the mean-field MPCD-MD Debye-H\"{u}ckel algorithm used here~\cite{hickey12}. 

Our results demonstrate that in microfluidic devices, frictional drag with walls is not entirely responsible for changing mobility --- wall-chain interactions increase drag competing with alignment and counterion condensation effects. 
We expect that our fundamental findings will aid the design of new electrophoretic methods for highly charged biomolecules. 
While the polyelectrolytes in this work are freely-jointed with a finite Debye length, DNA has a large persistence length. 
Radial confinement of DNA would still orient local segments; however, $N$-dependent charge condensation due to overlapping Debye layers is predicted to be less significant. 

$\qquad$\newline
This work was supported through an NSERC Discovery Grant to G.W.S and EMBO funding to T.N.S. (ALTF181-2013). 
Computational resources were provided by Sharcnet.

\bibliography{citations}

\begin{thebibliography}{42}%
\makeatletter
\providecommand \@ifxundefined [1]{%
 \@ifx{#1\undefined}
}%
\providecommand \@ifnum [1]{%
 \ifnum #1\expandafter \@firstoftwo
 \else \expandafter \@secondoftwo
 \fi
}%
\providecommand \@ifx [1]{%
 \ifx #1\expandafter \@firstoftwo
 \else \expandafter \@secondoftwo
 \fi
}%
\providecommand \natexlab [1]{#1}%
\providecommand \enquote  [1]{``#1''}%
\providecommand \bibnamefont  [1]{#1}%
\providecommand \bibfnamefont [1]{#1}%
\providecommand \citenamefont [1]{#1}%
\providecommand \href@noop [0]{\@secondoftwo}%
\providecommand \href [0]{\begingroup \@sanitize@url \@href}%
\providecommand \@href[1]{\@@startlink{#1}\@@href}%
\providecommand \@@href[1]{\endgroup#1\@@endlink}%
\providecommand \@sanitize@url [0]{\catcode `\\12\catcode `\$12\catcode
  `\&12\catcode `\#12\catcode `\^12\catcode `\_12\catcode `\%12\relax}%
\providecommand \@@startlink[1]{}%
\providecommand \@@endlink[0]{}%
\providecommand \url  [0]{\begingroup\@sanitize@url \@url }%
\providecommand \@url [1]{\endgroup\@href {#1}{\urlprefix }}%
\providecommand \urlprefix  [0]{URL }%
\providecommand \Eprint [0]{\href }%
\providecommand \doibase [0]{http://dx.doi.org/}%
\providecommand \selectlanguage [0]{\@gobble}%
\providecommand \bibinfo  [0]{\@secondoftwo}%
\providecommand \bibfield  [0]{\@secondoftwo}%
\providecommand \translation [1]{[#1]}%
\providecommand \BibitemOpen [0]{}%
\providecommand \bibitemStop [0]{}%
\providecommand \bibitemNoStop [0]{.\EOS\space}%
\providecommand \EOS [0]{\spacefactor3000\relax}%
\providecommand \BibitemShut  [1]{\csname bibitem#1\endcsname}%
\let\auto@bib@innerbib\@empty
\bibitem [{\citenamefont {Dorfman}\ \emph {et~al.}(2013)\citenamefont
  {Dorfman}, \citenamefont {King}, \citenamefont {Olson}, \citenamefont
  {Thomas},\ and\ \citenamefont {Tree}}]{dorfman13}%
  \BibitemOpen
  \bibfield  {author} {\bibinfo {author} {\bibfnamefont {K.~D.}\ \bibnamefont
  {Dorfman}}, \bibinfo {author} {\bibfnamefont {S.~B.}\ \bibnamefont {King}},
  \bibinfo {author} {\bibfnamefont {D.~W.}\ \bibnamefont {Olson}}, \bibinfo
  {author} {\bibfnamefont {J.~D.~P.}\ \bibnamefont {Thomas}}, \ and\ \bibinfo
  {author} {\bibfnamefont {D.~R.}\ \bibnamefont {Tree}},\ }\href@noop {}
  {\bibfield  {journal} {\bibinfo  {journal} {Chem. Rev.}\ }\textbf {\bibinfo
  {volume} {113}},\ \bibinfo {pages} {2584} (\bibinfo {year}
  {2013})}\BibitemShut {NoStop}%
\bibitem [{\citenamefont {de~Haan}\ and\ \citenamefont
  {Slater}(2013)}]{dehaan13}%
  \BibitemOpen
  \bibfield  {author} {\bibinfo {author} {\bibfnamefont {H.~W.}\ \bibnamefont
  {de~Haan}}\ and\ \bibinfo {author} {\bibfnamefont {G.~W.}\ \bibnamefont
  {Slater}},\ }\href@noop {} {\bibfield  {journal} {\bibinfo  {journal} {Phys.
  Rev. Lett.}\ }\textbf {\bibinfo {volume} {110}},\ \bibinfo {pages} {048101}
  (\bibinfo {year} {2013})}\BibitemShut {NoStop}%
\bibitem [{\citenamefont {Laohakunakorn}\ \emph {et~al.}(2013)\citenamefont
  {Laohakunakorn}, \citenamefont {Ghosal}, \citenamefont {Otto}, \citenamefont
  {Misiunas},\ and\ \citenamefont {Keyser}}]{laohakunakorn13}%
  \BibitemOpen
  \bibfield  {author} {\bibinfo {author} {\bibfnamefont {N.}~\bibnamefont
  {Laohakunakorn}}, \bibinfo {author} {\bibfnamefont {S.}~\bibnamefont
  {Ghosal}}, \bibinfo {author} {\bibfnamefont {O.}~\bibnamefont {Otto}},
  \bibinfo {author} {\bibfnamefont {K.}~\bibnamefont {Misiunas}}, \ and\
  \bibinfo {author} {\bibfnamefont {U.~F.}\ \bibnamefont {Keyser}},\
  }\href@noop {} {\bibfield  {journal} {\bibinfo  {journal} {Nano Lett.}\
  }\textbf {\bibinfo {volume} {13}},\ \bibinfo {pages} {2798} (\bibinfo {year}
  {2013})}\BibitemShut {NoStop}%
\bibitem [{\citenamefont {Rowghanian}\ and\ \citenamefont
  {Grosberg}(2013)}]{rowghanian13}%
  \BibitemOpen
  \bibfield  {author} {\bibinfo {author} {\bibfnamefont {P.}~\bibnamefont
  {Rowghanian}}\ and\ \bibinfo {author} {\bibfnamefont {A.~Y.}\ \bibnamefont
  {Grosberg}},\ }\href@noop {} {\bibfield  {journal} {\bibinfo  {journal}
  {Phys. Rev. E}\ }\textbf {\bibinfo {volume} {87}},\ \bibinfo {pages} {042723}
  (\bibinfo {year} {2013})}\BibitemShut {NoStop}%
\bibitem [{\citenamefont {Randall}\ and\ \citenamefont
  {Doyle}(2004)}]{randall04}%
  \BibitemOpen
  \bibfield  {author} {\bibinfo {author} {\bibfnamefont {G.}~\bibnamefont
  {Randall}}\ and\ \bibinfo {author} {\bibfnamefont {P.}~\bibnamefont
  {Doyle}},\ }\href@noop {} {\bibfield  {journal} {\bibinfo  {journal} {Phys.
  Rev. Lett.}\ }\textbf {\bibinfo {volume} {93}},\ \bibinfo {pages} {058102}
  (\bibinfo {year} {2004})}\BibitemShut {NoStop}%
\bibitem [{\citenamefont {Randall}\ and\ \citenamefont
  {Doyle}(2006)}]{randall06}%
  \BibitemOpen
  \bibfield  {author} {\bibinfo {author} {\bibfnamefont {G.}~\bibnamefont
  {Randall}}\ and\ \bibinfo {author} {\bibfnamefont {P.}~\bibnamefont
  {Doyle}},\ }\href@noop {} {\bibfield  {journal} {\bibinfo  {journal}
  {Macromolecules}\ }\textbf {\bibinfo {volume} {39}},\ \bibinfo {pages} {7734}
  (\bibinfo {year} {2006})}\BibitemShut {NoStop}%
\bibitem [{\citenamefont {Dorfman}(2010)}]{dorfman10}%
  \BibitemOpen
  \bibfield  {author} {\bibinfo {author} {\bibfnamefont {K.}~\bibnamefont
  {Dorfman}},\ }\href@noop {} {\bibfield  {journal} {\bibinfo  {journal} {Rev.
  Mod. Phys.}\ }\textbf {\bibinfo {volume} {82}},\ \bibinfo {pages} {2903}
  (\bibinfo {year} {2010})}\BibitemShut {NoStop}%
\bibitem [{\citenamefont {Viero}\ \emph {et~al.}(2011)\citenamefont {Viero},
  \citenamefont {He},\ and\ \citenamefont {Bancaud}}]{viero11}%
  \BibitemOpen
  \bibfield  {author} {\bibinfo {author} {\bibfnamefont {Y.}~\bibnamefont
  {Viero}}, \bibinfo {author} {\bibfnamefont {Q.}~\bibnamefont {He}}, \ and\
  \bibinfo {author} {\bibfnamefont {A.}~\bibnamefont {Bancaud}},\ }\href@noop
  {} {\bibfield  {journal} {\bibinfo  {journal} {Small}\ }\textbf {\bibinfo
  {volume} {7}},\ \bibinfo {pages} {3508} (\bibinfo {year} {2011})}\BibitemShut
  {NoStop}%
\bibitem [{\citenamefont {Park}\ \emph {et~al.}(2012)\citenamefont {Park},
  \citenamefont {Olson},\ and\ \citenamefont {Dorfman}}]{park12}%
  \BibitemOpen
  \bibfield  {author} {\bibinfo {author} {\bibfnamefont {S.-G.}\ \bibnamefont
  {Park}}, \bibinfo {author} {\bibfnamefont {D.~W.}\ \bibnamefont {Olson}}, \
  and\ \bibinfo {author} {\bibfnamefont {K.~D.}\ \bibnamefont {Dorfman}},\
  }\href@noop {} {\bibfield  {journal} {\bibinfo  {journal} {Lab Chip}\
  }\textbf {\bibinfo {volume} {12}},\ \bibinfo {pages} {1463} (\bibinfo {year}
  {2012})}\BibitemShut {NoStop}%
\bibitem [{\citenamefont {Mannion}\ \emph {et~al.}(2006)\citenamefont
  {Mannion}, \citenamefont {Reccius}, \citenamefont {Cross},\ and\
  \citenamefont {Craighead}}]{mannion2006cas}%
  \BibitemOpen
  \bibfield  {author} {\bibinfo {author} {\bibfnamefont {J.~T.}\ \bibnamefont
  {Mannion}}, \bibinfo {author} {\bibfnamefont {C.~H.}\ \bibnamefont
  {Reccius}}, \bibinfo {author} {\bibfnamefont {J.~D.}\ \bibnamefont {Cross}},
  \ and\ \bibinfo {author} {\bibfnamefont {H.~G.}\ \bibnamefont {Craighead}},\
  }\href@noop {} {\bibfield  {journal} {\bibinfo  {journal} {Biophys. J.}\
  }\textbf {\bibinfo {volume} {90}},\ \bibinfo {pages} {4538} (\bibinfo {year}
  {2006})}\BibitemShut {NoStop}%
\bibitem [{\citenamefont {Reisner}\ \emph {et~al.}(2005)\citenamefont
  {Reisner}, \citenamefont {Morton}, \citenamefont {Riehn}, \citenamefont
  {Wang}, \citenamefont {Yu}, \citenamefont {Rosen}, \citenamefont {Sturm},
  \citenamefont {Chou}, \citenamefont {Frey},\ and\ \citenamefont
  {Austin}}]{reisner2005sad}%
  \BibitemOpen
  \bibfield  {author} {\bibinfo {author} {\bibfnamefont {W.}~\bibnamefont
  {Reisner}}, \bibinfo {author} {\bibfnamefont {K.}~\bibnamefont {Morton}},
  \bibinfo {author} {\bibfnamefont {R.}~\bibnamefont {Riehn}}, \bibinfo
  {author} {\bibfnamefont {Y.}~\bibnamefont {Wang}}, \bibinfo {author}
  {\bibfnamefont {Z.}~\bibnamefont {Yu}}, \bibinfo {author} {\bibfnamefont
  {M.}~\bibnamefont {Rosen}}, \bibinfo {author} {\bibfnamefont
  {J.}~\bibnamefont {Sturm}}, \bibinfo {author} {\bibfnamefont
  {S.}~\bibnamefont {Chou}}, \bibinfo {author} {\bibfnamefont {E.}~\bibnamefont
  {Frey}}, \ and\ \bibinfo {author} {\bibfnamefont {R.}~\bibnamefont
  {Austin}},\ }\href@noop {} {\bibfield  {journal} {\bibinfo  {journal} {Phys.
  Rev. Lett.}\ }\textbf {\bibinfo {volume} {94}},\ \bibinfo {pages} {196101}
  (\bibinfo {year} {2005})}\BibitemShut {NoStop}%
\bibitem [{\citenamefont {Shendruk}\ \emph {et~al.}(2012)\citenamefont
  {Shendruk}, \citenamefont {Hickey}, \citenamefont {Slater},\ and\
  \citenamefont {Harden}}]{shendruk12}%
  \BibitemOpen
  \bibfield  {author} {\bibinfo {author} {\bibfnamefont {T.~N.}\ \bibnamefont
  {Shendruk}}, \bibinfo {author} {\bibfnamefont {O.~A.}\ \bibnamefont
  {Hickey}}, \bibinfo {author} {\bibfnamefont {G.~W.}\ \bibnamefont {Slater}},
  \ and\ \bibinfo {author} {\bibfnamefont {J.~L.}\ \bibnamefont {Harden}},\
  }\href@noop {} {\bibfield  {journal} {\bibinfo  {journal} {Curr. Opin.
  Colloid Interface Sci.}\ }\textbf {\bibinfo {volume} {17}},\ \bibinfo {pages}
  {74 } (\bibinfo {year} {2012})}\BibitemShut {NoStop}%
\bibitem [{\citenamefont {Long}\ and\ \citenamefont {Ajdari}(2001)}]{long01}%
  \BibitemOpen
  \bibfield  {author} {\bibinfo {author} {\bibfnamefont {D.}~\bibnamefont
  {Long}}\ and\ \bibinfo {author} {\bibfnamefont {A.}~\bibnamefont {Ajdari}},\
  }\href@noop {} {\bibfield  {journal} {\bibinfo  {journal} {Eur. Phys. J. E}\
  }\textbf {\bibinfo {volume} {4}},\ \bibinfo {pages} {29} (\bibinfo {year}
  {2001})}\BibitemShut {NoStop}%
\bibitem [{\citenamefont {Viovy}(2000)}]{viovy00}%
  \BibitemOpen
  \bibfield  {author} {\bibinfo {author} {\bibfnamefont {J.}~\bibnamefont
  {Viovy}},\ }\href@noop {} {\bibfield  {journal} {\bibinfo  {journal} {Rev.
  Mod. Phys.}\ }\textbf {\bibinfo {volume} {72}},\ \bibinfo {pages} {813}
  (\bibinfo {year} {2000})}\BibitemShut {NoStop}%
\bibitem [{\citenamefont {Desruisseaux}\ \emph {et~al.}(2001)\citenamefont
  {Desruisseaux}, \citenamefont {Long}, \citenamefont {Drouin},\ and\
  \citenamefont {Slater}}]{desruisseaux01}%
  \BibitemOpen
  \bibfield  {author} {\bibinfo {author} {\bibfnamefont {C.}~\bibnamefont
  {Desruisseaux}}, \bibinfo {author} {\bibfnamefont {D.}~\bibnamefont {Long}},
  \bibinfo {author} {\bibfnamefont {G.}~\bibnamefont {Drouin}}, \ and\ \bibinfo
  {author} {\bibfnamefont {G.~W.}\ \bibnamefont {Slater}},\ }\href@noop {}
  {\bibfield  {journal} {\bibinfo  {journal} {Macromolecules}\ }\textbf
  {\bibinfo {volume} {34}},\ \bibinfo {pages} {44} (\bibinfo {year}
  {2001})}\BibitemShut {NoStop}%
\bibitem [{\citenamefont {Long}\ \emph {et~al.}(1996)\citenamefont {Long},
  \citenamefont {Viovy},\ and\ \citenamefont {Ajdari}}]{long1996sae}%
  \BibitemOpen
  \bibfield  {author} {\bibinfo {author} {\bibfnamefont {D.}~\bibnamefont
  {Long}}, \bibinfo {author} {\bibfnamefont {J.}~\bibnamefont {Viovy}}, \ and\
  \bibinfo {author} {\bibfnamefont {A.}~\bibnamefont {Ajdari}},\ }\href@noop {}
  {\bibfield  {journal} {\bibinfo  {journal} {Phys. Rev. Lett.}\ }\textbf
  {\bibinfo {volume} {76}},\ \bibinfo {pages} {3858} (\bibinfo {year}
  {1996})}\BibitemShut {NoStop}%
\bibitem [{\citenamefont {Chubynsky}\ and\ \citenamefont
  {Slater}(2014)}]{chubynsky14}%
  \BibitemOpen
  \bibfield  {author} {\bibinfo {author} {\bibfnamefont {M.~V.}\ \bibnamefont
  {Chubynsky}}\ and\ \bibinfo {author} {\bibfnamefont {G.~W.}\ \bibnamefont
  {Slater}},\ }\href@noop {} {\bibfield  {journal} {\bibinfo  {journal}
  {Electrophoresis}\ }\textbf {\bibinfo {volume} {35}},\ \bibinfo {pages} {596}
  (\bibinfo {year} {2014})}\BibitemShut {NoStop}%
\bibitem [{\citenamefont {Tlusty}(2006)}]{tlusty06}%
  \BibitemOpen
  \bibfield  {author} {\bibinfo {author} {\bibfnamefont {T.}~\bibnamefont
  {Tlusty}},\ }\href@noop {} {\bibfield  {journal} {\bibinfo  {journal}
  {Macromolecules}\ }\textbf {\bibinfo {volume} {39}},\ \bibinfo {pages} {3927}
  (\bibinfo {year} {2006})}\BibitemShut {NoStop}%
\bibitem [{\citenamefont {Balducci}\ \emph {et~al.}(2006)\citenamefont
  {Balducci}, \citenamefont {Mao}, \citenamefont {Han},\ and\ \citenamefont
  {Doyle}}]{balducci06}%
  \BibitemOpen
  \bibfield  {author} {\bibinfo {author} {\bibfnamefont {A.}~\bibnamefont
  {Balducci}}, \bibinfo {author} {\bibfnamefont {P.}~\bibnamefont {Mao}},
  \bibinfo {author} {\bibfnamefont {J.}~\bibnamefont {Han}}, \ and\ \bibinfo
  {author} {\bibfnamefont {P.}~\bibnamefont {Doyle}},\ }\href@noop {}
  {\bibfield  {journal} {\bibinfo  {journal} {Macromolecules}\ }\textbf
  {\bibinfo {volume} {39}},\ \bibinfo {pages} {6273} (\bibinfo {year}
  {2006})}\BibitemShut {NoStop}%
\bibitem [{\citenamefont {Salieb-Beugelaar}\ \emph {et~al.}(2008)\citenamefont
  {Salieb-Beugelaar}, \citenamefont {Teapal}, \citenamefont {van
  Nieuwkasteele}, \citenamefont {Wijnperl\'{e}}, \citenamefont {Tegenfeldt},
  \citenamefont {Lisdat}, \citenamefont {van~den Berg},\ and\ \citenamefont
  {Eijkel}}]{scalieb08}%
  \BibitemOpen
  \bibfield  {author} {\bibinfo {author} {\bibfnamefont {G.}~\bibnamefont
  {Salieb-Beugelaar}}, \bibinfo {author} {\bibfnamefont {J.}~\bibnamefont
  {Teapal}}, \bibinfo {author} {\bibfnamefont {J.}~\bibnamefont {van
  Nieuwkasteele}}, \bibinfo {author} {\bibfnamefont {D.}~\bibnamefont
  {Wijnperl\'{e}}}, \bibinfo {author} {\bibfnamefont {J.}~\bibnamefont
  {Tegenfeldt}}, \bibinfo {author} {\bibfnamefont {F.}~\bibnamefont {Lisdat}},
  \bibinfo {author} {\bibfnamefont {A.}~\bibnamefont {van~den Berg}}, \ and\
  \bibinfo {author} {\bibfnamefont {J.}~\bibnamefont {Eijkel}},\ }\href@noop {}
  {\bibfield  {journal} {\bibinfo  {journal} {Nano Lett.}\ }\textbf {\bibinfo
  {volume} {8}},\ \bibinfo {pages} {1785} (\bibinfo {year} {2008})}\BibitemShut
  {NoStop}%
\bibitem [{\citenamefont {Cross}\ \emph {et~al.}(2007)\citenamefont {Cross},
  \citenamefont {Strychalski},\ and\ \citenamefont {Craighead}}]{cross07}%
  \BibitemOpen
  \bibfield  {author} {\bibinfo {author} {\bibfnamefont {J.}~\bibnamefont
  {Cross}}, \bibinfo {author} {\bibfnamefont {E.}~\bibnamefont {Strychalski}},
  \ and\ \bibinfo {author} {\bibfnamefont {H.}~\bibnamefont {Craighead}},\
  }\href@noop {} {\bibfield  {journal} {\bibinfo  {journal} {J. Appl. Phys.}\
  }\textbf {\bibinfo {volume} {102}},\ \bibinfo {pages} {024701} (\bibinfo
  {year} {2007})}\BibitemShut {NoStop}%
\bibitem [{\citenamefont {Campbell}\ \emph {et~al.}(2004)\citenamefont
  {Campbell}, \citenamefont {Wilkinson}, \citenamefont {Manz}, \citenamefont
  {Camilleri},\ and\ \citenamefont {Humphreys}}]{campbell04}%
  \BibitemOpen
  \bibfield  {author} {\bibinfo {author} {\bibfnamefont {L.~C.}\ \bibnamefont
  {Campbell}}, \bibinfo {author} {\bibfnamefont {M.~J.}\ \bibnamefont
  {Wilkinson}}, \bibinfo {author} {\bibfnamefont {A.}~\bibnamefont {Manz}},
  \bibinfo {author} {\bibfnamefont {P.}~\bibnamefont {Camilleri}}, \ and\
  \bibinfo {author} {\bibfnamefont {C.~J.}\ \bibnamefont {Humphreys}},\
  }\href@noop {} {\bibfield  {journal} {\bibinfo  {journal} {Lab Chip}\
  }\textbf {\bibinfo {volume} {4}},\ \bibinfo {pages} {225} (\bibinfo {year}
  {2004})}\BibitemShut {NoStop}%
\bibitem [{\citenamefont {Salieb-Beugelaar}(2009)}]{salieb2009electrokinetic}%
  \BibitemOpen
  \bibfield  {author} {\bibinfo {author} {\bibfnamefont {G.~B.}\ \bibnamefont
  {Salieb-Beugelaar}},\ }\emph {\bibinfo {title} {Electrokinetic transport of
  DNA in nanoslits}},\ \href@noop {} {Ph.D. thesis},\ \bibinfo  {school}
  {University of Twente} (\bibinfo {year} {2009})\BibitemShut {NoStop}%
\bibitem [{\citenamefont {Kang}\ \emph {et~al.}(2006)\citenamefont {Kang},
  \citenamefont {Lee},\ and\ \citenamefont {Yeung}}]{kang2006e}%
  \BibitemOpen
  \bibfield  {author} {\bibinfo {author} {\bibfnamefont {S.~H.}\ \bibnamefont
  {Kang}}, \bibinfo {author} {\bibfnamefont {S.}~\bibnamefont {Lee}}, \ and\
  \bibinfo {author} {\bibfnamefont {E.~S.}\ \bibnamefont {Yeung}},\ }\href@noop
  {} {\bibfield  {journal} {\bibinfo  {journal} {Electrophoresis}\ }\textbf
  {\bibinfo {volume} {27}},\ \bibinfo {pages} {4149} (\bibinfo {year}
  {2006})}\BibitemShut {NoStop}%
\bibitem [{\citenamefont {Math\'{e}}\ \emph {et~al.}(2007)\citenamefont
  {Math\'{e}}, \citenamefont {Meglio},\ and\ \citenamefont
  {Tinland}}]{mathe07}%
  \BibitemOpen
  \bibfield  {author} {\bibinfo {author} {\bibfnamefont {J.}~\bibnamefont
  {Math\'{e}}}, \bibinfo {author} {\bibfnamefont {J.-M.~D.}\ \bibnamefont
  {Meglio}}, \ and\ \bibinfo {author} {\bibfnamefont {B.}~\bibnamefont
  {Tinland}},\ }\href@noop {} {\bibfield  {journal} {\bibinfo  {journal} {J.
  Colloid Interf. Sci.}\ }\textbf {\bibinfo {volume} {316}},\ \bibinfo {pages}
  {831 } (\bibinfo {year} {2007})}\BibitemShut {NoStop}%
\bibitem [{\citenamefont {Dai}\ \emph {et~al.}(2013)\citenamefont {Dai},
  \citenamefont {Tree}, \citenamefont {van~der Maarel}, \citenamefont
  {Dorfman},\ and\ \citenamefont {Doyle}}]{dai13}%
  \BibitemOpen
  \bibfield  {author} {\bibinfo {author} {\bibfnamefont {L.}~\bibnamefont
  {Dai}}, \bibinfo {author} {\bibfnamefont {D.~R.}\ \bibnamefont {Tree}},
  \bibinfo {author} {\bibfnamefont {J.~R.~C.}\ \bibnamefont {van~der Maarel}},
  \bibinfo {author} {\bibfnamefont {K.~D.}\ \bibnamefont {Dorfman}}, \ and\
  \bibinfo {author} {\bibfnamefont {P.~S.}\ \bibnamefont {Doyle}},\ }\href@noop
  {} {\bibfield  {journal} {\bibinfo  {journal} {Phys. Rev. Lett.}\ }\textbf
  {\bibinfo {volume} {110}},\ \bibinfo {pages} {168105} (\bibinfo {year}
  {2013})}\BibitemShut {NoStop}%
\bibitem [{\citenamefont {Long}\ and\ \citenamefont
  {Ajdari}(1996)}]{long1996emc}%
  \BibitemOpen
  \bibfield  {author} {\bibinfo {author} {\bibfnamefont {D.}~\bibnamefont
  {Long}}\ and\ \bibinfo {author} {\bibfnamefont {A.}~\bibnamefont {Ajdari}},\
  }\href@noop {} {\bibfield  {journal} {\bibinfo  {journal} {Electrophoresis}\
  }\textbf {\bibinfo {volume} {17}},\ \bibinfo {pages} {1161} (\bibinfo {year}
  {1996})}\BibitemShut {NoStop}%
\bibitem [{\citenamefont {Grosberg}\ and\ \citenamefont
  {Rabin}(2010)}]{grosberg10}%
  \BibitemOpen
  \bibfield  {author} {\bibinfo {author} {\bibfnamefont {A.~Y.}\ \bibnamefont
  {Grosberg}}\ and\ \bibinfo {author} {\bibfnamefont {Y.}~\bibnamefont
  {Rabin}},\ }\href@noop {} {\bibfield  {journal} {\bibinfo  {journal} {J.
  Chem. Phys.}\ }\textbf {\bibinfo {volume} {133}},\ \bibinfo {eid} {165102}
  (\bibinfo {year} {2010})}\BibitemShut {NoStop}%
\bibitem [{\citenamefont {Hickey}\ \emph {et~al.}(2012)\citenamefont {Hickey},
  \citenamefont {Shendruk}, \citenamefont {Harden},\ and\ \citenamefont
  {Slater}}]{hickey12}%
  \BibitemOpen
  \bibfield  {author} {\bibinfo {author} {\bibfnamefont {O.}~\bibnamefont
  {Hickey}}, \bibinfo {author} {\bibfnamefont {T.}~\bibnamefont {Shendruk}},
  \bibinfo {author} {\bibfnamefont {J.}~\bibnamefont {Harden}}, \ and\ \bibinfo
  {author} {\bibfnamefont {G.}~\bibnamefont {Slater}},\ }\href@noop {}
  {\bibfield  {journal} {\bibinfo  {journal} {Phys. Rev. Lett.}\ } (\bibinfo
  {year} {2012})}\BibitemShut {NoStop}%
\bibitem [{\citenamefont {Slater}\ \emph {et~al.}(2009)\citenamefont {Slater},
  \citenamefont {Holm}, \citenamefont {Chubynsky}, \citenamefont {de~Haan},
  \citenamefont {Dub\'{e}}, \citenamefont {Grass}, \citenamefont {Hickey},
  \citenamefont {Kingsburry}, \citenamefont {Sean}, \citenamefont {Shendruk},\
  and\ \citenamefont {Zhan}}]{slater09}%
  \BibitemOpen
  \bibfield  {author} {\bibinfo {author} {\bibfnamefont {G.~W.}\ \bibnamefont
  {Slater}}, \bibinfo {author} {\bibfnamefont {C.}~\bibnamefont {Holm}},
  \bibinfo {author} {\bibfnamefont {M.~V.}\ \bibnamefont {Chubynsky}}, \bibinfo
  {author} {\bibfnamefont {H.~W.}\ \bibnamefont {de~Haan}}, \bibinfo {author}
  {\bibfnamefont {A.}~\bibnamefont {Dub\'{e}}}, \bibinfo {author}
  {\bibfnamefont {K.}~\bibnamefont {Grass}}, \bibinfo {author} {\bibfnamefont
  {O.~A.}\ \bibnamefont {Hickey}}, \bibinfo {author} {\bibfnamefont
  {C.}~\bibnamefont {Kingsburry}}, \bibinfo {author} {\bibfnamefont
  {D.}~\bibnamefont {Sean}}, \bibinfo {author} {\bibfnamefont {T.~N.}\
  \bibnamefont {Shendruk}}, \ and\ \bibinfo {author} {\bibfnamefont
  {L.}~\bibnamefont {Zhan}},\ }\href@noop {} {\bibfield  {journal} {\bibinfo
  {journal} {Electrophoresis}\ }\textbf {\bibinfo {volume} {30}},\ \bibinfo
  {pages} {792} (\bibinfo {year} {2009})}\BibitemShut {NoStop}%
\bibitem [{\citenamefont {Jiang}\ \emph {et~al.}(2013)\citenamefont {Jiang},
  \citenamefont {Watari},\ and\ \citenamefont {Larson}}]{jiang13}%
  \BibitemOpen
  \bibfield  {author} {\bibinfo {author} {\bibfnamefont {L.}~\bibnamefont
  {Jiang}}, \bibinfo {author} {\bibfnamefont {N.}~\bibnamefont {Watari}}, \
  and\ \bibinfo {author} {\bibfnamefont {R.~G.}\ \bibnamefont {Larson}},\
  }\href@noop {} {\bibfield  {journal} {\bibinfo  {journal} {J. Rheol.}\
  }\textbf {\bibinfo {volume} {57}},\ \bibinfo {pages} {1177} (\bibinfo {year}
  {2013})}\BibitemShut {NoStop}%
\bibitem [{\citenamefont {Noguchi}\ \emph {et~al.}(2007)\citenamefont
  {Noguchi}, \citenamefont {Kikuchi},\ and\ \citenamefont
  {Gompper}}]{noguchi07}%
  \BibitemOpen
  \bibfield  {author} {\bibinfo {author} {\bibfnamefont {H.}~\bibnamefont
  {Noguchi}}, \bibinfo {author} {\bibfnamefont {N.}~\bibnamefont {Kikuchi}}, \
  and\ \bibinfo {author} {\bibfnamefont {G.}~\bibnamefont {Gompper}},\
  }\href@noop {} {\bibfield  {journal} {\bibinfo  {journal} {Europhys. Lett.}\
  }\textbf {\bibinfo {volume} {78}},\ \bibinfo {pages} {10005} (\bibinfo {year}
  {2007})}\BibitemShut {NoStop}%
\bibitem [{\citenamefont {Frank}\ and\ \citenamefont
  {Winkler}(2009)}]{frank09}%
  \BibitemOpen
  \bibfield  {author} {\bibinfo {author} {\bibfnamefont {S.}~\bibnamefont
  {Frank}}\ and\ \bibinfo {author} {\bibfnamefont {R.~G.}\ \bibnamefont
  {Winkler}},\ }\href@noop {} {\bibfield  {journal} {\bibinfo  {journal} {J.
  Chem. Phys.}\ }\textbf {\bibinfo {volume} {131}} (\bibinfo {year}
  {2009})}\BibitemShut {NoStop}%
\bibitem [{\citenamefont {Padding}\ and\ \citenamefont
  {Louis}(2004)}]{padding04}%
  \BibitemOpen
  \bibfield  {author} {\bibinfo {author} {\bibfnamefont {J.~T.}\ \bibnamefont
  {Padding}}\ and\ \bibinfo {author} {\bibfnamefont {A.~A.}\ \bibnamefont
  {Louis}},\ }\href@noop {} {\bibfield  {journal} {\bibinfo  {journal} {Phys.
  Rev. Lett.}\ }\textbf {\bibinfo {volume} {93}},\ \bibinfo {pages} {220601}
  (\bibinfo {year} {2004})}\BibitemShut {NoStop}%
\bibitem [{\citenamefont {Dai}\ and\ \citenamefont {Doyle}(2013)}]{dai13b}%
  \BibitemOpen
  \bibfield  {author} {\bibinfo {author} {\bibfnamefont {L.}~\bibnamefont
  {Dai}}\ and\ \bibinfo {author} {\bibfnamefont {P.~S.}\ \bibnamefont
  {Doyle}},\ }\href@noop {} {\bibfield  {journal} {\bibinfo  {journal}
  {Macromolecules}\ }\textbf {\bibinfo {volume} {46}},\ \bibinfo {pages} {6336}
  (\bibinfo {year} {2013})}\BibitemShut {NoStop}%
\bibitem [{\citenamefont {Shew}(2003)}]{shew2003cbs}%
  \BibitemOpen
  \bibfield  {author} {\bibinfo {author} {\bibfnamefont {C.}~\bibnamefont
  {Shew}},\ }\href@noop {} {\bibfield  {journal} {\bibinfo  {journal} {J. Chem.
  Phys.}\ }\textbf {\bibinfo {volume} {119}},\ \bibinfo {pages} {10428}
  (\bibinfo {year} {2003})}\BibitemShut {NoStop}%
\bibitem [{\citenamefont {Kikuchi}\ \emph {et~al.}(2002)\citenamefont
  {Kikuchi}, \citenamefont {Gent},\ and\ \citenamefont {Yeomans}}]{kikuchi02}%
  \BibitemOpen
  \bibfield  {author} {\bibinfo {author} {\bibfnamefont {N.}~\bibnamefont
  {Kikuchi}}, \bibinfo {author} {\bibfnamefont {A.}~\bibnamefont {Gent}}, \
  and\ \bibinfo {author} {\bibfnamefont {J.}~\bibnamefont {Yeomans}},\
  }\href@noop {} {\bibfield  {journal} {\bibinfo  {journal} {Euro. Phys. J. E}\
  }\textbf {\bibinfo {volume} {9}},\ \bibinfo {pages} {63} (\bibinfo {year}
  {2002})}\BibitemShut {NoStop}%
\bibitem [{\citenamefont {Kikuchi}\ \emph {et~al.}(2005)\citenamefont
  {Kikuchi}, \citenamefont {Ryder}, \citenamefont {Pooley},\ and\ \citenamefont
  {Yeomans}}]{kikuchi05}%
  \BibitemOpen
  \bibfield  {author} {\bibinfo {author} {\bibfnamefont {N.}~\bibnamefont
  {Kikuchi}}, \bibinfo {author} {\bibfnamefont {J.~F.}\ \bibnamefont {Ryder}},
  \bibinfo {author} {\bibfnamefont {C.~M.}\ \bibnamefont {Pooley}}, \ and\
  \bibinfo {author} {\bibfnamefont {J.~M.}\ \bibnamefont {Yeomans}},\
  }\href@noop {} {\bibfield  {journal} {\bibinfo  {journal} {Phys. Rev. E}\
  }\textbf {\bibinfo {volume} {71}},\ \bibinfo {pages} {061804} (\bibinfo
  {year} {2005})}\BibitemShut {NoStop}%
\bibitem [{\citenamefont {Micheletti}\ \emph {et~al.}(2011)\citenamefont
  {Micheletti}, \citenamefont {Marenduzzo},\ and\ \citenamefont
  {Orlandini}}]{micheletti11}%
  \BibitemOpen
  \bibfield  {author} {\bibinfo {author} {\bibfnamefont {C.}~\bibnamefont
  {Micheletti}}, \bibinfo {author} {\bibfnamefont {D.}~\bibnamefont
  {Marenduzzo}}, \ and\ \bibinfo {author} {\bibfnamefont {E.}~\bibnamefont
  {Orlandini}},\ }\href@noop {} {\bibfield  {journal} {\bibinfo  {journal}
  {Phys. Rep.}\ }\textbf {\bibinfo {volume} {504}},\ \bibinfo {pages} {1 }
  (\bibinfo {year} {2011})}\BibitemShut {NoStop}%
\bibitem [{\citenamefont {Muthukumar}(2004)}]{muthukumar04}%
  \BibitemOpen
  \bibfield  {author} {\bibinfo {author} {\bibfnamefont {M.}~\bibnamefont
  {Muthukumar}},\ }\href@noop {} {\bibfield  {journal} {\bibinfo  {journal} {J.
  Chem. Phys.}\ }\textbf {\bibinfo {volume} {120}} (\bibinfo {year}
  {2004})}\BibitemShut {NoStop}%
\bibitem [{\citenamefont {Morrison}\ and\ \citenamefont
  {Thirumalai}(2005)}]{morrison05}%
  \BibitemOpen
  \bibfield  {author} {\bibinfo {author} {\bibfnamefont {G.}~\bibnamefont
  {Morrison}}\ and\ \bibinfo {author} {\bibfnamefont {D.}~\bibnamefont
  {Thirumalai}},\ }\href@noop {} {\bibfield  {journal} {\bibinfo  {journal} {J.
  Chem. Phys.}\ }\textbf {\bibinfo {volume} {122}} (\bibinfo {year}
  {2005})}\BibitemShut {NoStop}%
\bibitem [{\citenamefont {Werner}\ \emph {et~al.}(2013)\citenamefont {Werner},
  \citenamefont {Westerlund}, \citenamefont {Tegenfeldt},\ and\ \citenamefont
  {Mehlig}}]{werner13}%
  \BibitemOpen
  \bibfield  {author} {\bibinfo {author} {\bibfnamefont {E.}~\bibnamefont
  {Werner}}, \bibinfo {author} {\bibfnamefont {F.}~\bibnamefont {Westerlund}},
  \bibinfo {author} {\bibfnamefont {J.~O.}\ \bibnamefont {Tegenfeldt}}, \ and\
  \bibinfo {author} {\bibfnamefont {B.}~\bibnamefont {Mehlig}},\ }\href@noop {}
  {\bibfield  {journal} {\bibinfo  {journal} {Macromolecules}\ }\textbf
  {\bibinfo {volume} {46}},\ \bibinfo {pages} {6644} (\bibinfo {year}
  {2013})}\BibitemShut {NoStop}%
\end{thebibliography}%

\end{document}